\documentclass[graybox, envcountchap]{svmult}

\usepackage{mathptmx}        
\usepackage{amsmath}
\usepackage{amssymb}
\usepackage{color}
\usepackage{helvet}          
\usepackage{courier}         
\usepackage{dirtree}

\usepackage{makeidx}        
\usepackage{graphicx}        
                                           
\usepackage{subfig}

\usepackage{multicol}        
\usepackage[bottom]{footmisc}

\usepackage{hyperref}      
\hypersetup{colorlinks=true,urlcolor=blue}

\usepackage[misc]{ifsym}

\makeindex

\begin{document}

\title{Photometric Redshifts}
\author{Luca Tortorelli and Daniel Gr\"un}
\institute{Luca Tortorelli (\Letter) \at University Observatory, Faculty of Physics, Ludwig-Maximilians-Universität, Scheinerstr. 1, 81679 Munich, Germany, \email{luca.tortorelli@lmu.de}
\and Daniel Gr\"un (\Letter) \at University Observatory, Faculty of Physics, Ludwig-Maximilians-Universität, Scheinerstr. 1, 81679 Munich, Germany, \email{daniel.gruen@lmu.de}}
\maketitle
\abstract{
The cosmological redshift of a galaxy's light is inferable from its observable properties in images. Because imaging is much easier to acquire than spectroscopic observations that would allow the identification of distinct line features, this motivates the technique of photometric redshift estimation (photo-$z$). Photo-$z$ has been an early and sustained driver for the utilization of artificial intelligence (AI) in astrophysics, and conversely AI methods are underlying most of the recent advances in photo-$z$. Here we review the diversity of AI methods applied to the photo-$z$ problem over the years in a discriminative way, that is, to regress redshift from photometric observables. We argue that, besides optimization suiting specific applications, this approach has effectively converged. It is limited not by the AI methodology but by the size and substantial systematic uncertainties and selection effects in spectroscopic training samples. In order to progress, either an unobtainable quantity and quality of training data or a more principled approach in using it is required. We thus outline ongoing research of integrating AI in a  Bayesian modeling of galaxy data. This comes in the form of generative models for representing the distribution of intrinsic properties and outcomes of telescope observations of the galaxy population.
}

\section{Introduction}

Estimating distances to objects in the Universe is one of the most fundamental problems in astronomy. Indeed, almost every inference about a celestial object starts from a distance: estimating luminosity from flux requires a luminosity distance, estimating physical from angular size an angular diameter distance, determining volumes and number densities requires comoving distances, and look-back time (another type of distance) places an observed galaxy on the chart of cosmic history. 

The distance of an object is a quantity whose measurement from an image is only possible through detours. Distances of celestial objects beyond our Milky Way need to be determined through a chain of methods that form the so-called "distance ladder": starting from trigonometric parallaxes, where one infers distance from the apparent annual motion of a nearby star against the background sky, every sub-sequent step of the ladder, which relies on standard candles such as RR Lyrae, Cepheids, Blue Supergiant stars, or type Ia supernovae, is calibrated upon the previous one to reach more and more distant objects, at the cost of inheriting systematics from the previous step and adding its own set of astrophysical assumptions. 

\subsubsection*{Redshift as a distance measure}

For large cosmological distances, the workhorse observable is the redshift. The stretching of the wavelength of photons that originate from objects on extragalactic distances provides a direct measurement on the expansion of the Universe's scale factor during their travel time. This so-called redshift can be related to the distance to the light source (when assuming a cosmological model) by means of the cosmic expansion law.  Distinguishing galaxies by their redshift hence allows us to track the evolution of the galaxy population and the growth of large scale structures over cosmic time, ultimately testing our picture of galaxy formation and of the physics governing the Universe as a whole.

The most precise way to obtain redshifts for extra-galactic sources is via spectroscopy. Using a dispersing element in the light path of a telescope, one can resolve a galaxy’s light as a function of wavelength to identify spectral features (emission or absorption lines and breaks). The wavelength ratios between pairs of these features provide a secure identification of the features. The ratio of the observed wavelength to the known wavelength of the feature at rest then allows a redshift determination. In modern galaxy redshift surveys, spectroscopic resolving powers $R= \lambda / \Delta \lambda$ range from $R\sim 10^2$ to $R \gtrsim 10^4$. These enable redshift uncertainties far below $R^{-1}$ when signal-to-noise is sufficient and multiple features are detected. The central limitation is the cost of spectroscopic observations. In order to identify a spectroscopic feature, it needs to be detectable against the noise. In addition to just the photon counting noise in the intrinsically small amount of light corresponding to a spectral feature, the signal from an astronomical source gets dispersed over a wider pixel area on the detector whose readout noise variances add. As a result, spectroscopy demands substantially increased exposure time compared to photometric observations just to avoid being dominated by this noise component. Spectroscopic observations are further constrained by finite multiplexing, meaning that, for the most feasible techniques, only a small number (tens to few thousands) of fibers or slits can be put on astronomical targets. This problem is enormously exacerbated by the scale of next-generation imaging surveys, such as the Vera C. Rubin Observatory’s Legacy Survey of Space and Time (hereafter, LSST, \cite{Ivezic2019}) or the Euclid space telescope\cite{Mellier2025}. At the expected density of the galaxy samples to be well enough measured for scientific application in their final data, a single LSST frame contains roughly a million galaxies. Even with highly multiplexed instruments, such as the Dark Energy Spectroscopic Instrument (DESI, \cite{DESI}), 4MOST \cite{DeJong2019}, and the Prime Focus Spectrograph (PFS, \cite{PFS}), obtaining spectroscopic redshifts for the billions of galaxies composing these “gold samples” of Stage IV surveys \cite{Albrecht2006} is not realistic, as estimated e.g.~in\cite{NewmanGruen2022}. This is the fundamental motivation for photometric redshift estimation: it is the only scalable option to obtain distances for the faint and numerous galaxies that dominate the statistical power of modern cosmology and much of galaxy evolution.

\subsubsection*{Photometric redshift estimates}

Photometric redshifts (or, photo-$z$) infer redshift from (commonly broad-band) fluxes that can be measured in a sky image, rather than from spectra. Operationally, a photo-$z$ method exploits the relation that exists between a galaxy's redshift and its apparent colors, the so-called color-redshift relation. This relation is due to the fact that characteristic features in the spectral energy distribution of a galaxy are mapped to different photometric filters depending on an object's redshift. Photo-$z$ methods (see \cite{Salvato2019} for a review) use observed photometry to provide either a point estimate $\hat{z}$ of the cosmological redshift $z$ or an estimate $\hat{p}(z)$ of the full redshift (posterior) probability density $p(z)$, i.e.
\begin{equation}
\mathrm{photo-}z: \lbrace \mathrm{photometric\; measurements} \rbrace \rightarrow \hat{z} \; ,
\end{equation}
or 
\begin{equation}
\mathrm{photo-}z: \lbrace \mathrm{photometric\; measurements} \rbrace \rightarrow \hat{p}(z) \; .
\end{equation}

The details of what is included in the photometric measurements and how the mapping is implemented leads to a great diversity of approaches. This includes whether and how artificial intelligence (hereafter, AI) is used in the course, but also how the result is interpreted, e.g. as an estimator with certain desired or undesired properties or as a Bayesian posterior $\hat{p}(z)=p(z | \mathrm{photometric\; measurements}, \mathrm{model})$ with some explicit or implicit model assumptions about galaxies and the measurement process. 

Different scientific applications lead to different requirements on the quantities estimated in the above equations. For instance, there may be a threshold $\delta_z$ for the allowable bias of the photo-$z$ estimate, i.e.~
\begin{equation}
|\mathbb{E}(\hat{z}-z)|< \delta_z
\end{equation}
or
\begin{equation}
\left|z-\int_z dz \, z \, \hat{p}(z)\right|<\delta_z \; .
\end{equation}
Similar requirements may exist on higher moments of the redshift (error) distribution, e.g.~for the true redshift to be within the bulk of $\hat{p}(z)$ or within some number of standard deviations of the error of $\hat{z}$ some fraction of the time.

\subsubsection*{Limitations to photo-$z$}

\begin{figure}
    \centering
    \includegraphics[width=1\linewidth]{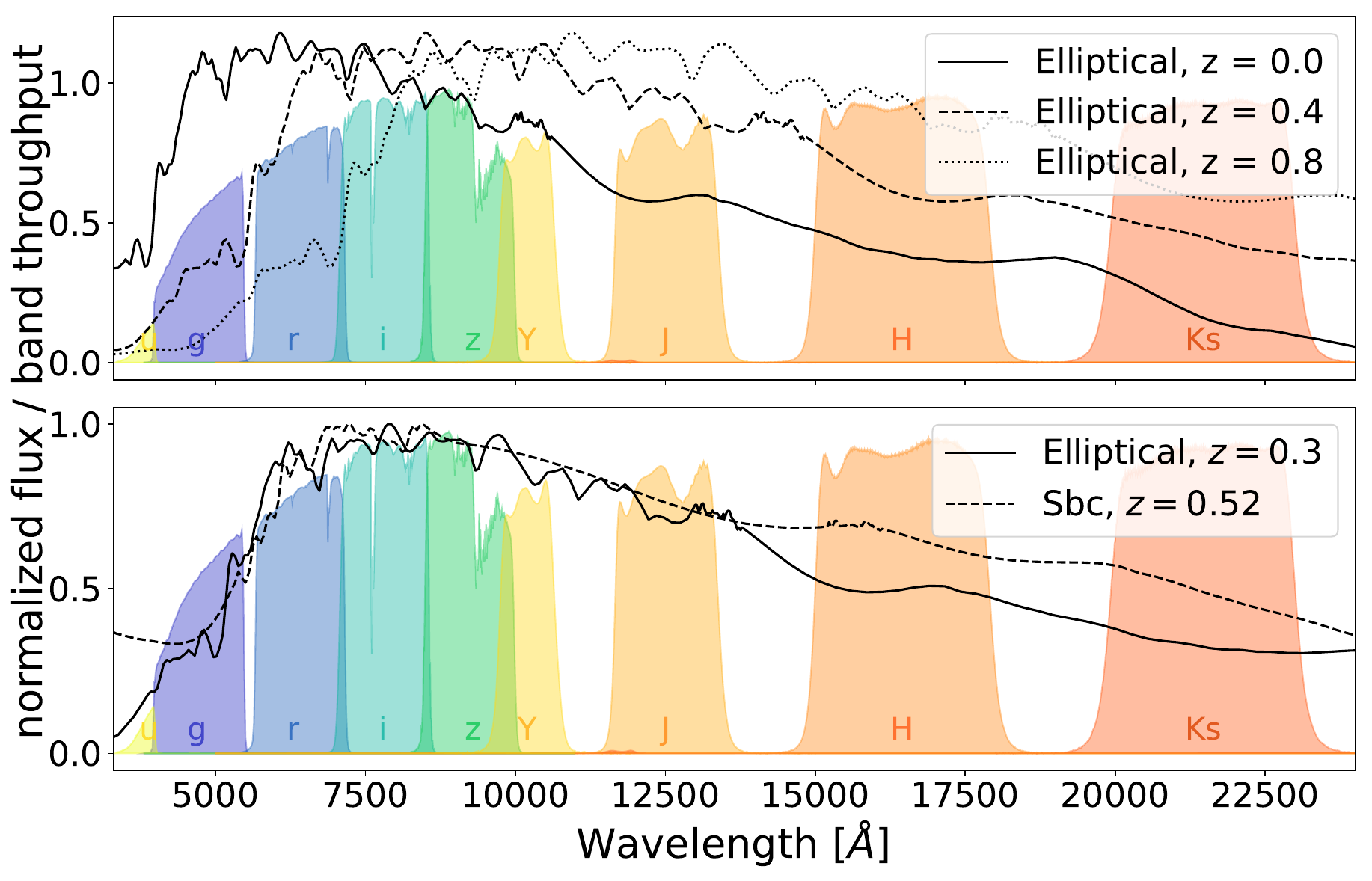}
    \caption{Figure 1 of \cite{Buchs2019} illustrating both the promise and the curse of photo-$z$ estimation. For sufficient photometric information and a limited diversity of intrinsic spectral energy distributions, redshift can be uniquely determined from color. This is the case in the top panel, showing passive galaxies at different redshift whose $4000 \ \mathrm{\AA}$ break moves across broad-band filters. Type-redshift degeneracies (bottom panel) arise when only few optical broad-band filters are observed (e.g. $griz$) that are indistinguishable for different intrinsic spectral energy distributions (SEDs) at different redshift and can thus lead to multi-modal $p(z)$ and "catastrophic outliers". The addition of UV and near-IR photometry in the example shown and in general greatly helps in breaking this degeneracy. Colored areas refer to the throughput of Dark Energy Survey (DES, \cite{DES}) optical $ugriz$ and Visible and Infrared Survey Telescope for Astronomy (VISTA, \cite{VISTA}) $YJHK_s$ infrared bands.
    }
    \label{fig:tortorelli_gruen_2026_fig1}
\end{figure}

The observational limitation to the precision of photo-$z$ is spectral resolution which, if lacking, can mean that irreducible ambiguity remains. With broad-band filters, the effective resolution is $R<10$, meaning that only a small set of spectral features that are significant enough over such a wide wavelength range (strong breaks and overall slopes of the spectral energy distribution) can be used to constrain redshift. This leads to degeneracies between multiple ways to fit a galaxy's observed spectral energy distribution \cite{Buchs2019} due to, e.g., confusion between the $4000 \ \mathrm{\AA}$ break at lower redshift and the Lyman break at higher redshift. Such "type-redshift degeneracy" (Fig. \ref{fig:tortorelli_gruen_2026_fig1}) leads to multi-modal $p(z)$ and to "catastrophic outliers" where $|\hat{z} - z|$ is large. These effects are worse when photometric observations only exist in few filters and when the signal-to-noise ratios of the photometric data are low.

A more addressable limitation lies in not having a good enough knowledge of the galaxy population to set an accurate prior on the redshift of a photometrically observed galaxy. That is often a dominant concern because the resulting impact on the photo-$z$ estimate does not average out over a sample of galaxies. This issue is mirrored in shortcomings of any photo-$z$ method, even ones that do not attempt to explicitly include a model of the galaxy population in the process. The limitation is related to the type-redshift degeneracy described above: a knowledge of the possible rest-frame spectral energy distributions is required to determine at what observed color there is a degeneracy, how it could be broken, and, in the absence of data that would allow that, what mix of types and redshifts a set of photometrically indistinguishable galaxies are likely to have. Even if photometry were perfect, the mapping from colors to redshift depends on how galaxies populate color space as a function of redshift, type, dust, metallicity, and emission line properties. Any mismatch between the assumed population model (e.g.~a template set with associated priors) or the training sample (e.g.~via a spectroscopic selection) and the true galaxy population in the target survey will bias the photo-$z$ estimation. This is true for individual galaxy $p(z)$s but, due to the required accuracy, particularly at the level of the redshift distribution $N(z)$ of a photometrically selected ensemble of galaxies (see \cite{NewmanGruen2022} for a comprehensive review). This bias and limited performance propagate into any scientific applications of photo-$z$.

\subsubsection*{Applications of photo-$z$}

Photo-$z$s are fundamental for both galaxy evolution and cosmological studies. Galaxy evolution aims to determine  how individual galaxies and their populations change with redshift: stellar mass assembly, star formation quenching, morphological transformation, the emergence of the bimodality of galaxy types, and environmental impact. Photo-$z$s enable these studies because they unlock large, deep, and homogeneous samples. This is in contrast to spectroscopic samples for galaxy evolution that exist mostly in pencil-beamed (e.g. DEVILS \cite{Davies2018}, VIPERS \cite{Guzzo2014}) or bright, low-redshift surveys (e.g. GAMA \cite{Driver2022}). However, photo-$z$ uncertainties propagate into essentially every derived quantity. Redshift errors bias luminosities, rest-frame colors, spectral types, and therefore stellar masses and star formation rates derived from photometry. Even when analyses use broad redshift bins, catastrophic outliers can inject e.g.~low-redshift interlopers into high-redshift samples of intrinsically bright galaxies, systematically distorting inferred evolutionary trends. Environmental studies are also impacted as overdensity-finding algorithms \cite{Rykoff2014,Oguri2014,Bellagamba2018,Aguena2021} effectively count neighbors in redshift cylinders. These are smeared along the line of sight by photo-$z$ scatter, thereby increasing chance projections and degrading the signal-to-noise ratio. Photo-$z$ outliers suppress clustering amplitudes and bias inferred halo occupation interpretations unless their rate is known and accounted for. 

Cosmology, in turn, uses galaxies as tracers of the large scale structure growth. For this, redshifts or ensemble redshift distributions are always required. Weak gravitational lensing analyses \cite{Mandelbaum2018,Prat} typically bin source galaxies by redshift, making their respective $N(z)$ a first-order ingredient in predicting weak lensing shear observables. Galaxy clustering and $3 \times 2$pt  analyses (using shear-shear, galaxy-galaxy lensing, and galaxy clustering correlation functions) in addition depend on the $N(z)$ of a "lens" galaxy sample. This encodes how likely two galaxies near one another in the sky are to also be near one another along the line of sight and hence physically clustered. The mapping between angular and physical scales, and the time evolution of clustering bias and structure growth further require knowing redshift. Uncertainties in the $N(z)$ hence represent one of the leading observational systematics in large scale structure cosmology. This is testified by a number of cosmic shear results from Stage III surveys that show how the degree of consistency of cosmological parameters between late and early Universe probes is severely affected by the photo-$z$ calibration strategies adopted. In the case of weak lensing, for instance, cosmic shear analyses from the Kilo-Degree Survey (KiDS-1000, \cite{Asgari2021}) and, to a lesser degree, Hyper Suprime-Cam (HSC) Year 3 (Y3) Results \cite{Dalal2023}, and DES Y3 \cite{Amon2022}, hinted at the existence of a tension with CMB data \cite{Planck2018}. 

This tension appeared in the values of the matter density $\Omega_{\rm m}$, the amplitude of matter density fluctuations $\sigma_8$, and in particular their combination $S_8 \equiv \sigma_8 (\Omega_{\rm m}/0.3)^{0.5}$, which traces the integrated growth of structure, and is known as `$S_8$ tension' \cite{Pantos2026}. Late Universe probes pointed towards an $S_8$ value of $S_8 \approx 0.75-0.78$, while the CMB constrains $S_8$ to be $S_8=0.832 \pm 0.013$ \cite{Planck2020}. The significance of this difference historically ranged from $1.5$ to over $3 \sigma$. This finding sparked a debate in the community, with proposed solutions ranging from "new" physics to observational systematics \cite{Amon2022b}. Two recent cosmic shear results seem to suggest that the origin of this tension lies in the latter. An improved redshift estimation method and calibration sample was the main change in the KiDS-Legacy cosmic shear analysis \cite{Wright2025} that led the collaboration to infer a value of $S_8=0.815^{+0.016}_{-0.021}$, with no hint of a tension with the CMB. Similarly, a re-analysis of HSC Y3 data using a new calibration of the tomographic redshift distributions via DESI clustering redshifts \cite{Janvry2025} shifted the $S_8$ value considerably higher towards Planck cosmology. The most precise such measurement to date is likewise not adding to the $S_8$ tension: the cosmological analysis from the full six years of observations of DES (DES Y6). It exhibits an estimate of $S_8$ that is $2.0 \sigma –2.6 \sigma$ below that of Planck from cosmic shear alone ($S_8= 0.783 \pm ^{+0.019}_{-0.015}$) and in the 3x2pt analysis ($S_8= 0.789 \pm 0.012$), respectively. Besides the impact of baryonic physics and intrinsic alignment \cite{Bigwood2026}, confident cosmological interpretation of any such parameter differences is limited primarily by the lack of representative spectroscopic redshifts of faint galaxies in the calibration sample.

Given these limitations, Stage IV experiments have proclaimed stringent requirements on the tolerated uncertainty in the mean redshift of each source galaxy bin. For LSST weak lensing analysis, the LSST Dark Energy Science Collaboration (DESC) science requirements \cite{LSSTScienceRequirements} sets targets at the per-mille level, $|\delta z| < 0.002 \times (1 + z)$ in the year 1 analysis, tightening to $|\delta z| < 0.001 \times (1 + z)$ in the year 10 analysis. Euclid similarly adopts target uncertainties on the mean redshift of  bins at the $0.002 \times (1 + z)$ level \cite{Euclid}. However, current photo-$z$ methods used in Stage III cosmology \cite{Rau2023, Wright2025b, Yin2025} still fall short of these goals by factors of $\approx 10$ due to uncertainties on the mean redshift that are degraded by spectroscopic incompleteness and outliers in the calibration redshifts, sample variance, blending, and astrophysical systematics \cite{NewmanGruen2022}. 

\subsubsection*{Artificial Intelligence to the rescue?}

In addition to the a-priori attractiveness and accessibility of the photo-$z$ problem for AI methods, this critical challenge motivates AI as a key to enabling insights about fundamental physics by resolving the redshift calibration gap. We here trace the diversity of ways in which this has been proposed and executed so far to answer  whether and how AI can indeed come to the rescue. 

Two classes of methods for photo-$z$ estimation have historically been distinguished which, for the purposes of this chapter, we will denote as \emph{discriminative} and \emph{generative} approaches. Historically the methods we categorize as discriminative have often been called "training based" and "empirical" versus generative "template fitting" approaches. As we will find there is a blurred line between these terms in particular from an AI perspective, e.g. insofar as templates can be based on empirical training, hence our choice of naming the classes. 

The discriminative class has traditionally been a field dominated by a great variety of machine learning techniques, with characteristic shortcomings of such approaches and the available training data setting limitations on their performance and characterization. The generative class has so far almost exclusively been based on human-led astrophysical modeling, although recently recognized limitations and approaches are increasingly leading to an adoption of AI here, too.    

\section{Discriminative AI for photometric redshifts}
\label{sec:1}

There is a sense in which "happy" photo-$z$ algorithms are all alike. They should return, for each galaxy, a Bayesian posterior 
\begin{equation}
\hat{p}(z) = \int \mathrm{d}t \, p(z, t | d) \propto \int \mathrm{d}t \, p(d | z, t) p(z, t) \; ,
\label{eqn:zbayes}
\end{equation}
where $d$ is all photometric information available on the galaxy, $t$ stands for a continuum of "types" of galaxies, $p(z, t| d)$ the joint posterior of type and redshift,  $p(d| z, t)$ the likelihood of the measurement process, and $p(z,t)$ a prior for the joint distribution of redshift and type. It is a matter of choice here whether $t$ is taken to be an observationally defined "phenotype" \cite{Sanchez2019} (in which case the likelihood is independent of $z$ and all the $z$ dependence is in the prior) or a rest-frame "genotype", in which case $z$ still impacts the expected observational signatures. In this notation, luminosity is subsumed in $t$, although it may in practice be useful to separate it out. Given the prior, which requires full understanding of galaxy evolution, and the likelihood, which encodes the full photometric measurement process, there is no ambiguity to this solution. Section~\ref{sec:generative} will describe ways in which AI may be of use in building (approximations to) such an ideal photo-$z$ posterior. 

Every "unhappy" photo-$z$ algorithm is "unhappy" in its own way though, and this may serve to categorize the discriminative approaches that can be applied in practice. We take discriminative methods in photo-$z$ to mean that an estimate $\hat{z}$ or $\hat{p}(z)$ is produced from the photometric measurements by some scheme trained on a sample of galaxies with true redshift labels and similar photometric measurements, i.e.

\begin{equation}
\mathrm{discriminative\; photo-}z: \lbrace \mathrm{photometric\; measurements\rbrace} \xrightarrow{{\substack{\text{training data,}\\\text{assumptions}}}} \hat{p}(z)\; \mathrm{or}\; \hat{z} \; .
\end{equation}

Each of the deviations these discriminative methods may show from the ideal estimator poses a certain trade off between e.g.~variance and bias, required observational, computational, or human resources. The metrics relevant for the utility of a photo-$z$ estimate, and hence the degree of unhappiness it may cause with the researcher relying on it, depend on the intended application. For this reason the correct way to strike that balance will differ from case to case. A diverse landscape of methods does and should exist, which we will attempt to describe from an AI perspective below.

\subsubsection*{From fitting a line to a diverse artificial intelligence methodology}

A first broad categorization of discriminative algorithms can be made between those that produce a discrete so-called point estimate $\hat{z}$ of an object's redshift against those that aim to produce a probability density function $\hat{p}(z)$. The former is really a special case of the latter since $\hat{z}$ is typically accommodated by metrics describing the error distribution, e.g.~a value for the typical standard deviation $\sigma_z$ of $\hat{z}$, from which an approximate Gaussian $\hat{p}(z)\propto\mathcal{N}(z-\hat{z},\sigma_z)$ could be formulated. Higher order information of the error distribution, commonly in the form of $N-\sigma$ outlier rates, is often also of relevance. These can be derived from the performance of the point estimate on training data with known true redshift, although with the caveat that they will depend on the sample, of which the training data may not be representative.

\begin{figure}
    \centering
    \includegraphics[width=\linewidth]{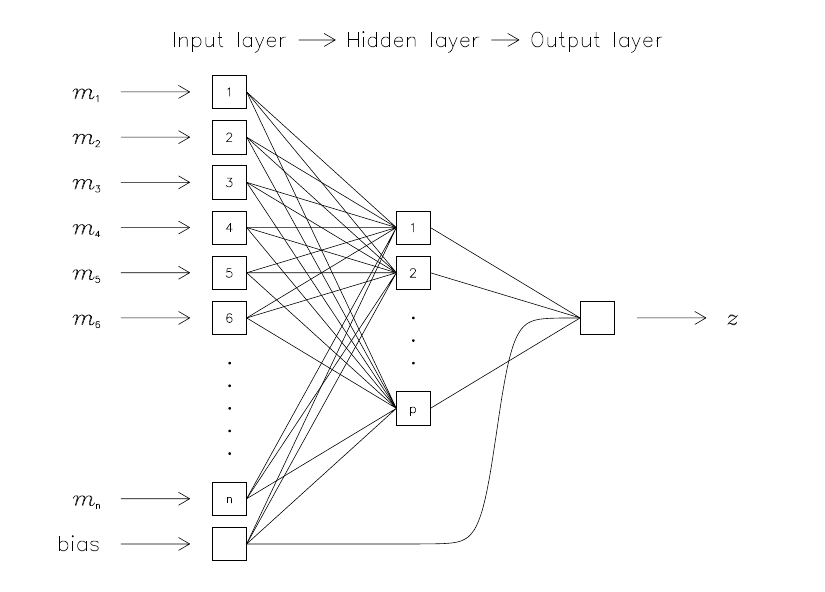}
    \caption{Photo-$z$ point estimation from observed magnitudes $m_i$ with a fully connected artificial neural network, introduced by ANNz \cite{Firth2003,Collister2004}, was a key starting point of AI in photo-$z$ and astrophysics more generally. Source: Figure 1 of \cite{Collister2004}, \copyright~The Astronomical Society of the Pacific. Reproduced by permission of IOP Publishing. All rights reserved.
    }
    \label{fig:tortorelli_gruen_2026_fig2}
\end{figure}

Early attempts at photo-$z$ estimation with empirical relations commonly were in the form of point estimates, with a continuous development of human-defined fitting functions into AI. \cite{Baum1962} determined a linear relation between the characteristic magnitude of cluster galaxies and the logarithm of their redshift. This first attempt was improved in a multi-linear or higher order fashion by e.g. \cite{Connolly1995,Brunner1997,Wang1998} with increasing complexity. 

The introduction of AI (beyond polynomial fitting) to photo-$z$ marks a significant milestone in both the evolution of photo-$z$ and the introduction of AI methods to astrophysics. Some of the early studies \cite{Firth2003,Tagliaferri2003,Collister2004} applying artificial neural networks that take photometric inputs to produce redshift point estimates as outputs (Figure~\ref{fig:tortorelli_gruen_2026_fig2}), trained using a sample of spectroscopically observed reference galaxies, remain impactful until today. Especially ANNz \cite{Firth2003,Collister2004} has been used as a reference implementation of multi-layer fully connected perceptrons (MLPs) in the astrophysics literature \cite{Collister2007, Gruen2010, Ross2011, Soumagnac2015}. 

The next stage of AI photo-$z$ research shows a proliferation of methods along multiple directions, each of which is one way of more closely approximating $p(z |$ all photometric information, model for galaxy population and observations$)$ than the $\lbrace \mathrm{magnitudes} \rbrace \rightarrow \hat{z}$ mapping by simple MLPs. Among the machine learning algorithms that have been applied to regress a redshift point prediction as a function of photometry, Gaussian Processes have been notably widely adopted \cite{Almosallam2016,Duncan2022}.

\subsubsection*{From point predictions to neural density estimates}

For many applications, a point estimate $\hat{z}$ is insufficient, however close to optimal it may be. Rather, the probability density of redshift of an individual galaxy on ensemble is the desired quantity.

In the neural network paradigm, point estimates can be generalized to parametric estimates of a redshift probability density (PDF). This happens e.g.~through mixture density networks that predict parameters of a functional form, such as a Gaussian mixture, of $\hat{p}(z)$ \cite{DIsanto2017}, or by means of probabilistically classifying galaxies as belonging to bins of a redshift histogram \cite{Rau2015}. Alternatively, ensembles of networks and/or perturbed input data can be  constructed to produce an approximate PDF \cite{Sadeh2016,Amaro2017,Cavuoti2017}. More flexible ways of estimating $\hat{p}(z)$ come from non-parametric methods, such as normalizing flows \cite{Crenshaw2024,Ren2026}. Here in effect a flexible PDF (in the form of a mapping of samples drawn from a standard normal distribution) is fitted to match the samples of spectroscopic redshifts in a training set, conditioned on observed photometric properties. It is not always clear that these methods encode the full uncertainty of the redshift estimate in the resulting $\hat{p}(z)$, which would have to contain e.g.~photometric errors in both training and application data, intrinsic ambiguity and additional uncertainty of the model of the galaxy color-magnitude-redshift relation, and limited sample size, model expressivity, and randomness during training.

\subsubsection*{From direct calibration to a bias-variance trade-off in clustering algorithms}

The diverse and complex AI methods in the previous chapter, perhaps surprisingly, only play a minor role in the state of the art of photo-$z$ for cosmological galaxy surveys. This is particularly true for those with a stringent requirement on the accuracy of the recovered redshift distributions. These most commonly ratherutilize a seemingly separate branch of AI algorithms that naturally produce redshift posterior PDFs: clustering algorithms.

To understand the suitability of clustering algorithms, first let us describe a frequentist picture for what is referred to by a redshift PDF. Consider the set of galaxies whose observed properties fit some definition of a class $c$, e.g.~by meeting some color-magnitude selection criteria. This could e.g.~be a redshift bin, or an even smaller subset of galaxies. If one was now able to spectroscopically observe a large number of such galaxies, representatively selected, ideally but impossibly over an infinite number of universes with equal physics but different random density fluctuations, then the histogram of those spectroscopic redshifts converges to the appropriate $p(z|c)$. While that limit is not obtainable in practice, one can get partway there in a variety of ways.

A relevant family of methods, subsumed here as $k$-nearest neighbors (kNN), uses the local environment of a given galaxy in photometric space to estimate a redshift or redshift distribution based on the subset of the training sample located inside of it. Kernel density estimates are one variant of this where the impact of a galaxy in the training sample is weighted in some way by its distance in photometric space from the galaxy of interest. This kind of approach is naturally matched to learning photo-$z$s and has been quite common in the past. Influential early such methods \cite{Lima2008,Ball2008,Cunha2009} have impacted later refinements. Notably, a number of flagship lensing analyses of the Stage III set of cosmological surveys have used elements of kNN for their redshift calibration \cite{Bonnett2016, Hildebrandt2017, Tanaka2018, Hoyle2018, Rau2023}. One downside of such approaches arises when the training data are sparse (which they always are when the photometric space is high-dimensional!) and hence the relevant neighbor(s) are far away in photometric space. Once the product of sample density and redshift varies non-linearly over that nearest-neighbor scale, a simple neighbor based estimate is generally biased. Directional Neighborhood Fitting is one response to this challenge that reduces the burden of bias-variance trade-off  and has found relatively widespread use \cite{DeVicente2016,Porredon2021,Weaverdyck2026}.

An equally simple approach not susceptible to this bias consists in (i) splitting the sample of relevant galaxies into subsets by their observed photometry, and (ii) interpreting the histogram of available spectroscopic or other high-quality redshift estimates of a subset $c$ as a noisy realization $\hat{p}(z|c)$ of $p(z|c)$.

A standard way of constructing such an estimator that clusters galaxies with similar redshift into the same $c$ are decision trees or, in a generalization that allows individual object $p(z)$ estimation, random forests. Historically forests were not behind neural networks for photo-$z$ estimation by long. Early works include \cite{Carliles2008,Gerdes2010}, the latter being a boosted decision tree method with superior performance on individual object $p(z)$ precision. Tree-based methods and later adaptations \cite{Carrasco2013,Gruen2014,Sadeh2016} have relevance in practice still today, although the use of clustering methods has diversified greatly, too. 

Other approaches of splitting the space of photometry exist, each with advantages and disadvantages relative to the above. Support vector machines (SVMs) have an ability to identify and use non-linear combinations of the input photometric properties to define galaxy subsets, a feature that is well-matched to the rugged structure of the color-redshift relation \cite{Wadadekar2005,Jones2017}. Works using two-dimensional self-organizing maps (SOMs) \cite{Masters2015,Buchs2019} have found wide adaptation by recent research in cosmology \cite{Masters2017,Wright2020,Myles2021,Yin2025} and galaxy evolution \cite{Davidzon2022,Teimoorinia2022,Holwerda2022,Abedini2026}, lately with the addition of the similar-in-spirit UMap clustering algorithm \cite{Ashmead2025}. While these dimensionality reduction techniques are not optimized to distinguish galaxies by redshift as well as photometrically possible, the notion of a low-dimensional manifold in high-dimensional color space matches the locus of galaxies well. The visualization with preserved neighborhoods in photometric space allows for interpretation and interpolation, which has added to the appeal of SOMs in photo-$z$.

Care must be taken in any of these clustering approaches that when the same redshift information is used for defining bins as for estimating their $N(z)$, Eddington-bias like effects are possible \cite{Cibirka2017,Roster2026}. These can be above the permissible level of systematic error and should be avoided through conscious bin definition.

\begin{figure}
    \centering
    \includegraphics[width=1\linewidth]{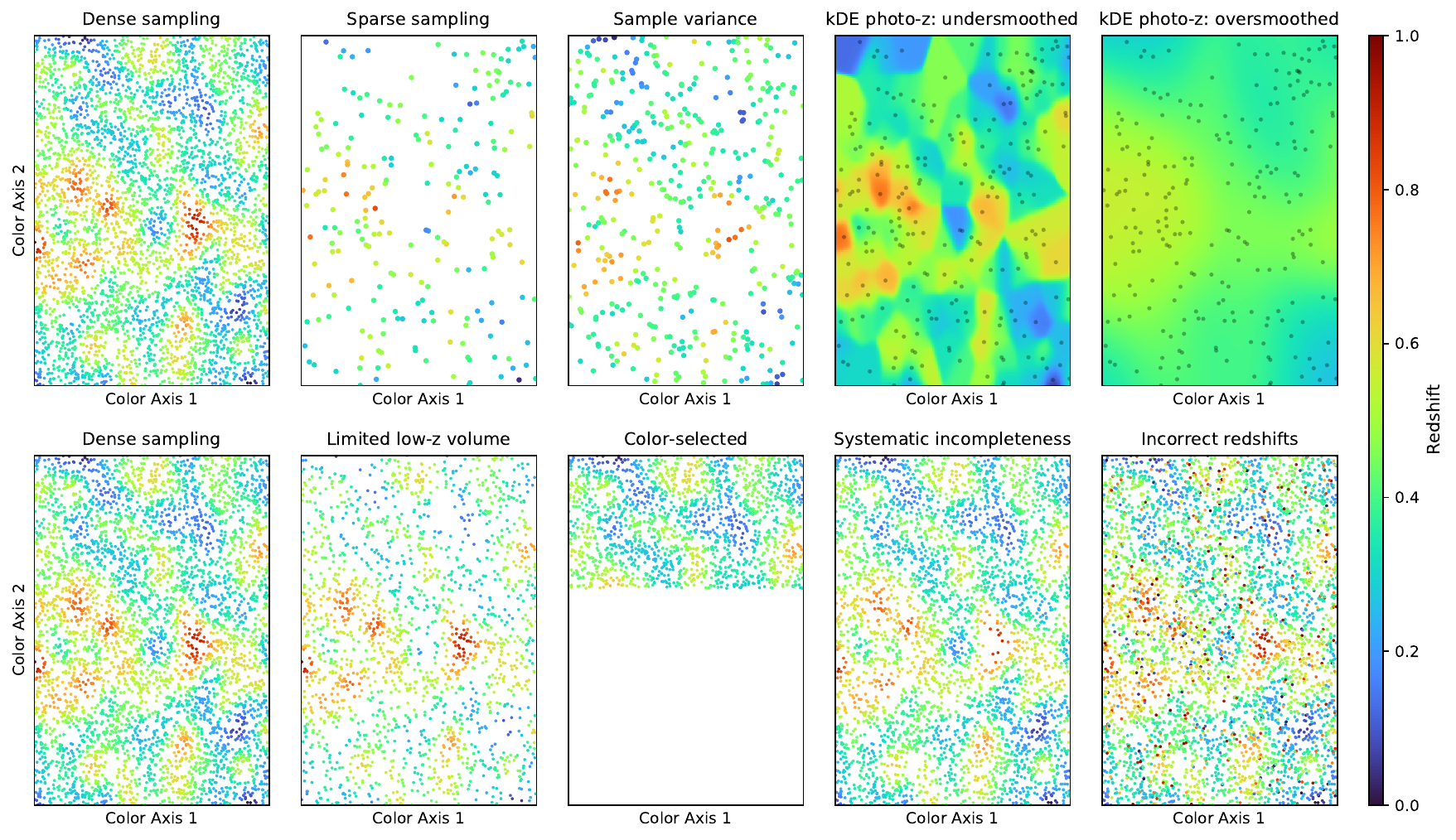}
    \caption{Illustration of the effects limiting clustering-based methods for photo-$z$, expanded from and created in a similar way as Figures~2 and 10 in \cite{NewmanGruen2022}. A dense sample of spectroscopic redshift measurements allows unbiased and minimum (although not zero) variance photo-$z$ estimation in every part of color space (first panel from left). Top panels show effects that are intrinsic to the requirement of a color neighborhood in these methods: the sparsity of training samples and the sample variance introduced by them being selected from small fields in the sky introduce ambiguity to that neighborhood (top, second and third panel). The required smoothing represents a trade-off between bias and variance (top, fourth and fifth panel). A variety of selection effects in spectroscopic training sample (bottom panels) further complicates the path to unbiased AI training.}
    \label{fig:tortorelli_gruen_2026_fig3}
\end{figure}

Each of these methods is in effect producing a redshift PDF estimate in the form of a weighted histogram of the redshifts of the training sample. A convenient feature of histograms is that the Bayesian treatment of their estimation is well understood. There exists a conjugate prior with respect to the likelihood of a histogram made from finite samples: the Dirichlet distribution \cite{Leistedt2016}. Hence, under the assumption that all galaxies in the training sample are independent of one another, uncertainty quantification in $\hat{p}(z)$ is possible in a principled fashion, even for multi-layer weighted versions of the histogram approach \cite{Sanchez2019, Sanchez2020, Alarcon2020}. 

In all such methods there is an intrinsic trade-off, by dialing up or down the smoothing scale, between the bias and the variance of the estimated $\hat{p}(z)$. When smoothing is done over large scales, e.g. subsets of galaxies chosen are large and not entirely indistinguishable by their observed photometric data, then $\hat{p}(z)$ will show an increased variance (cf. Figure~\ref{fig:tortorelli_gruen_2026_fig3}). However, in the limit of selecting the complete sample of galaxies an imaging experiment observes and creating an histogram for a representative sample of spectroscopic redshifts of them, the estimated $\hat{p}(z)$ will return an unbiased redshift distribution. This simple approach indeed leads the performance chart in most metrics given such a training sample (see the method \texttt{trainZ} in \cite{Schmidt2020}).

\subsubsection*{Unused information}

Equation~\ref{eqn:zbayes} mentioned "all photometric information", when most algorithms described above limit their input to flux measurements in a set of photometric bands, integrated over some aperture around the galaxy. One potential improvement therefore would be to instead truly use all photometric information available. Literature has explored the importance of various features \cite{Hoyle2015}, and some such features had been experimented with in the early MLP methods \cite{Firth2003}. The limiting case of using \emph{all} photometric information is well approximated by using galaxy images at the pixel level. The generalization of feed-forward MLP neural networks to take pixel-level image instead of catalog-level flux inputs was enabled by the use of convolutional networks or similar architectures \cite{Hoyle2016,Pasquet2019,Schuldt2021,Lin2022,Henghes2022,Dey2022,Treyer2024}. Indeed it has been found that for spatially resolved galaxy images, there is considerable additional information in the morphology available to the network from this kind of input. Related work went so far as to find that via a joint embedding of galaxy images and associated spectroscopic data (e.g. \cite{Parker2025}), detailed features of the spectra could be predicted from the images \cite{Doorenbos2024}. Notably, however, in deep ground-based imaging, the majority of galaxies is only barely resolved, greatly reducing the utility of images in addition to simply fluxes.

\subsubsection*{Implicit priors}

One way of interpreting the diversity of AI-based $\hat{p}$ in approximating Eq.~\ref{eqn:zbayes} is that implicitly the assumptions made in each method impose an imperfect prior on the observable-redshift relation of galaxies $p(z,t)$ (a point made by e.g. \cite{Schmidt2020}). Sometimes the distortion of the prior is straightforward to express. The example of \texttt{trainZ} mentioned above makes the assumption that $p(z,t)=p(z)=\int \mathrm{d}t p(z,t)$ is independent of the observable type $t$. More commonly, the assumptions made are more subtle, yet unless a methodology explicitly is constructed to infer and use $p(z,t)$ consistently, as no method discussed in this section does, it can be said to use an implicit prior.

\subsubsection*{Imperfect training samples}

To some degree a limiting factor to the methods described above are the assumptions made. At a more fundamental level, however, the limited information contained in a single galaxy in the training sample, and the limited number and imperfect representativeness of available and feasibly obtainable training samples introduces dominant uncertainty to photo-$z$ estimation with AI. These effects are not dissimilar to ones introducing biases in AI responses in other areas.

Full exploration of these effects goes beyond the scope of this review (see e.g. discussion and references in \cite{Gruen2017,NewmanGruen2022}), but the basic mechanisms are illustrated in the bottom panels of Figure~\ref{fig:tortorelli_gruen_2026_fig3}. In a given sky area, volume effects limit the number of low-redshift training galaxies available. Photometric pre-selection is part of almost all spectroscopic surveys and has to be accounted for in interpreting the recovered spectroscopic samples. A similar but more malign systematic incompleteness comes from unintended selection by spectroscopic redshift confidence. That is, at a given set of photometric observables of a galaxy, spectroscopic redshift measurement success is a function of a galaxy's true redshift, e.g.~because spectral features can move in or out of the wavelength range of the instrument. Finally, not all confident redshifts are correct. A failure mode particularly difficult to address is due to galaxies being blended together by random projection in the sky. These commonly get a single redshift assigned to them that is associated with the dominant spectral features, which may originate from a photometrically subdominant galaxy.

\subsubsection*{Ways forward}
\label{sect:ways_forward}

A way of mitigating the impact of both implicit priors and imperfect training samples lies in making the photometric subset definition of $c$ sharp enough. When $p(z|c)$ is strongly dominated by the likelihood, not the prior, and leaves little room in selection biases within the set of galaxies in $c$ to change the redshift by much, then the "happy" photo-$z$ estimator is closely approximated. This of course requires high quality photometry, and in addition suitable methodological choices, which has motivated the use of SOMs trained on multi-band deep photometry as part of the redshift estimation chain \cite{Masters2015,Buchs2019}.

The latest generations of the methods discussed in these sections, with the best training data sets available, still exceed the permissible photo-$z$ bias for next-generation experiments by about an order of magnitude. It does not appear feasible that gradual improvement in algorithms and spectroscopic training samples would reach the required fidelity in a few years time. It is for these reasons that at least for its highest stakes applications in cosmological studies of weak gravitational lensing and galaxy clustering, work on photo-$z$ estimation will have to start back at the drawing board, for which we outline a promising avenue in the following section.

\section{Generative AI for photometric redshifts}
\label{sec:generative}

We argued in the previous section that a "happy" photo-$z$ is a Bayesian posterior with the correct prior and likelihood, yet that no discriminative photo-$z$ method fully avoids the many ways in which one can deviate from that path. The alternative to discriminative photo-$z$ then is rather to forward model the process that generates the observations and learn from the comparison of the outcome of this and the real data. AI in this is used where needed to make the model sufficiently flexible and computationally tractable. In addition, AI techniques are typically needed for the inference step of the process. This change of paradigm can be seen as that instead of learning the inverse relation $\lbrace \mathrm{photometric \ measurement} \rbrace \rightarrow \hat{p}(z)$, one learns or emulates the ingredients that enter Bayes' theorem itself. 

Eqn. \ref{eqn:zbayes} had in a sense kept the model for what galaxies looked like, $p(z,t)$, and the likelihood as a description of the data taking process, $p(d|z,t)$, fixed. The uncertainty in both the galaxy population model and the observation model can instead also be made explicit in a hierarchical formulation. Both the prior and the likelihood then depend on additional population and observation hyper-parameters $\theta_{\mathrm{pop}}$ and $\theta_{\mathrm{obs}}$, such that the  joint posterior becomes
\begin{equation}p(z,t,\theta_{\mathrm{pop}},\theta_{\mathrm{obs}} | d) \propto p(d| z, t, \theta_{\mathrm{obs}}) p(z,t|\theta_{\mathrm{pop}}) p(\theta_{\mathrm{pop}},\theta_{\mathrm{obs}}) \; .
\label{eqn:bh}
\end{equation}

\begin{figure}
    \centering
    \includegraphics[width=1\linewidth]{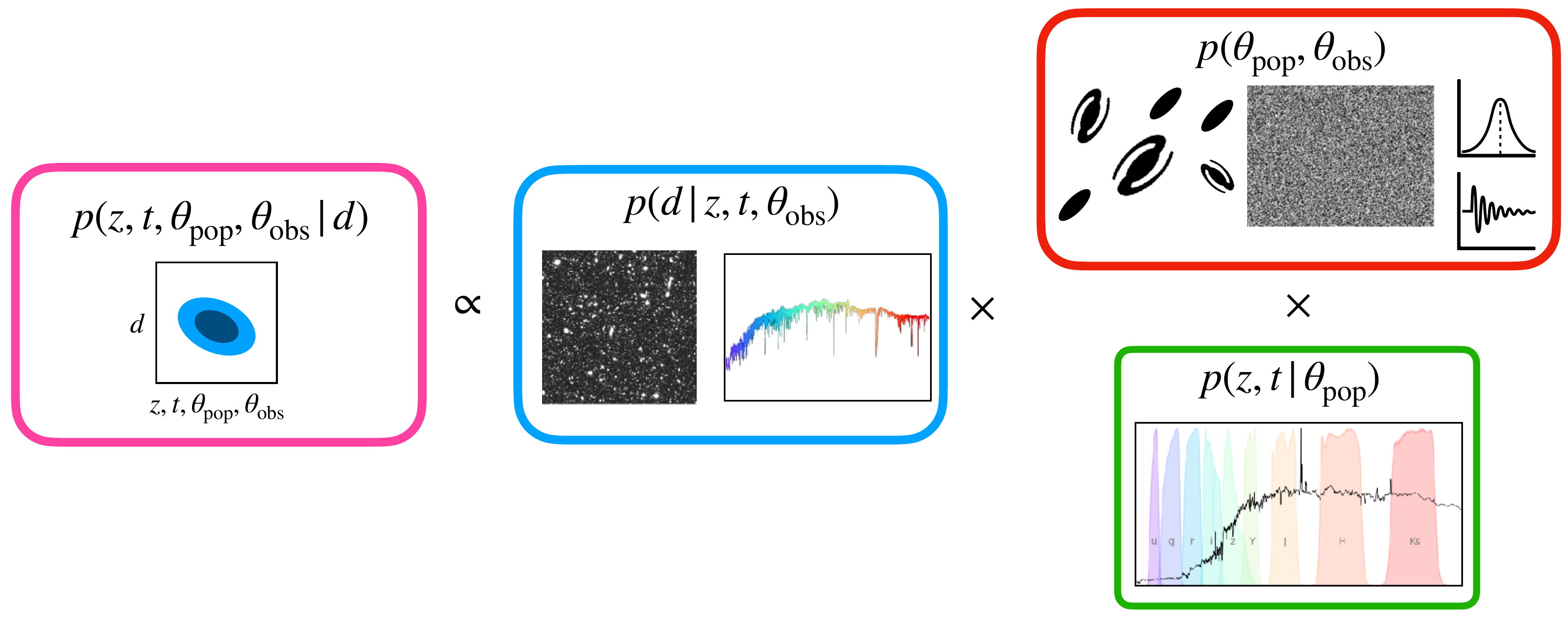}
    \caption{Illustration of the hierarchical formulation of photo-$z$ inference. The joint posterior of galaxy redshifts, types and population hyper-parameters (left, pink box) is obtained by modeling the prior distribution of galaxy population and observational and noise conditions parameters (top right, red box), the physical properties of galaxies conditioned on the population hyper-parameters (bottom right, green box), and the likelihood or data model that generates observed-like realizations given galaxy physical properties and observational and noise conditions (central, blue box). The simulated HSC image and the  galaxy SEDs are from the \textsc{galsbi} package \cite{Fischbacher2024,Fischbacher2025,Tortorelli2025}.}
    \label{fig:tortorelli_gruen_2026_fig4}
\end{figure}

This equation, illustrated in Fig. \ref{fig:tortorelli_gruen_2026_fig4}, makes clear that the joint posterior of galaxy redshifts, types, and population and observational hyper-parameters can and should be obtained in the context of an explicit forward model constrained by real data. Such a model must describe the intrinsic galaxy population through the hyper-parameters $\theta_{\mathrm{pop}}$, specify how latent galaxy properties map to redshifts and spectral types, and account for how observational conditions transform those properties into the measured photometric quantities. The latter term is sometimes known as data model. In generative approaches, both the prior and the likelihood are treated as explicit, constrainable components of the inference problem. The prior is no longer just a fixed choice of templates or an empirical calibration, but a flexible description of the galaxy population that can itself be constrained by the data. Likewise, the likelihood becomes an explicit model of the measurement process that can be calibrated and whose uncertainty can be propagated into downstream inferences. AI enters by providing flexible density models for the population term, fast emulators for expensive physical or observational calculations, and scalable inference engines for the resulting hierarchical posterior.

\subsubsection*{Approximated likelihood}
\label{sect:approximated_likelihood}

In the previous section we emphasized two shortcomings of discriminative methods, namely implicit priors and imperfect training samples. There is, however, a third way in which a photo-$z$ estimator can fail to be "happy": the likelihood itself can be wrong. In training-based methods without an explicit likelihood, this information is implicitly taken from the photometric measurements of the training sample, which are often not performed under the same conditions as for the sample of interest \cite{Hoyle2018,Lange2026}. When (e.g.~template fitting based) photo-$z$ methods explicitly evaluate a likelihood, images are usually first "compressed" into catalog-level fluxes and the likelihood is then approximated with a simple noise model around those summary measurements. This is approximately accurate for bright, isolated objects, but it becomes less precise exactly in those regimes that matter most for contemporary cosmology, meaning faint, blended, galaxies near the selection limit. To make matters worse, galaxies are observed with a noise level, point spread function (PSF), background, and deblending performance that vary as a function of sky position and wavelength. Under those circumstances, one cannot approximate the likelihood of a handful of observed catalog properties as having independent Gaussian errors. Rather, it is dominated by non-linear inter-connected effects that include detection, nonlinear model fitting, resulting measurement bias, blending, and selection, all of which couple photometric observables to redshift in a way that depends on the details of the survey and the galaxy population. 

A generative forward-modeling approach addresses this issue by treating the likelihood as a model for how latent galaxy properties, together with instrumental and observational conditions, give rise to the observed imaging data and the derived catalog measurements. It asks how likely the observed pixels, or the resulting catalog measurements, are given a galaxy with certain physical and morphological properties observed under specific instrumental and observational conditions. The ideal approach for this scenario is to forward model galaxy photons to images, run the same detection and measurement pipeline as the real survey, and compare data and simulations in the same space, for instance at the catalog level. This is the logic behind the development of pixel-level image simulations for forward-modeling approaches and selection function studies of wide-field galaxy surveys (see \cite{Plazas2020} for a review on image simulations specific for weak and strong gravitational lensing). These image simulators range from generating single or co-add survey images from user-defined input galaxy catalogs to inject synthetic sources into real images. Examples of the former include the \textsc{UFig} \cite{Berge2013,Fischbacher2025b} image simulator in the \textsc{GalSBI} framework \cite{Fischbacher2025,Tortorelli2025}, \textsc{SkyMaker} \cite{Bertin2009,Carassou2017,Schreiber2017}, \textsc{GalSim}-based \cite{Rowe2015} image simulators (e.g. LSST DESC DC2 and \textsc{ImSim}\footnote{https://github.com/LSSTDESC/Imsim} in the LSST survey) or survey-tailored simulators (e.g. for Euclid \cite{Serrano2024}). The latter category of synthetic source injectors includes \textsc{Balrog} \cite{Everett2022,Suchyta2016}, \textsc{Obiwan} \cite{Kong2020,Kong2025} and \textsc{SynPipe} \cite{Huang2018}, designed to evaluate the transfer function and study the imaging systematics of DES, the DESI Legacy Surveys, and HSC, respectively.

The challenge of this kind of image-level approaches is obvious: although being the most correct, realistic and robust approach, the pixel-level likelihood is expensive to evaluate. Some forward-modeling approaches, like \textsc{GalSBI}, mitigate this challenge by means of image simulators (\textsc{UFig}) that are as fast in rendering images as running common photometric extraction codes like \textsc{Source Extractor} \cite{Bertin1996} on them. However, even these optimized frameworks may require of order of millions of CPU hours for full image-level simulation-based inference on deep fields, which makes brute-force posterior exploration challenging for the large parameter spaces relevant to galaxy populations. 

This is where generative AI becomes useful in accelerating the mapping from the latent description of galaxies to their observed-like realizations. First attempts consisted in the use of variational autoencoders or generative adversarial networks to produce pixel-level simulations \cite{Smith2019,Lanusse2021,Bretonniere2022,Smith2022,Holzschuh2022,Campagne2025}. In addition to the faster rendering time, these approaches have also the advantage of being able to produce more complex morphologies than single or double S\'ersic \cite{Sersic1963} profiles, thanks to their training conducted on real galaxy images, and to the ability of condition these morphologies on the galaxy physical properties. More recent attempts focus instead on using neural emulators to directly map, at the catalog-level, the distribution of latent descriptions of galaxies to their photometrically measured properties, thereby replacing the entire pixel-level simulation. In \cite{Fischbacher2025} the authors used a normalizing flow emulator trained on \textsc{UFig} image simulations of HSC survey data to predict galaxy detection probabilities and measured observables, including the effects of blending, PSF variations, and background fluctuations, from \textsc{GalSBI} noiseless galaxy properties. Similarly, \cite{Leistedt2026} made use of the SURFS-based KiDS-Legacy-Like Simulations (\textsc{SKiLLS}, \cite{Li2023b}) to map \textsc{pop-cosmos} noiseless galaxy properties to KiDS-Legacy-like measured photometric quantities. Even the effect of galaxy images blending is accessible to such emulation \cite{Zhang2026}.

Once one accepts that the measurement process is part of the statistical model, pixel-level simulation and its emulation are no longer simply a more realistic way of doing the inference, but rather a central ingredient of the photo-$z$ problem, with AI necessary to address the scale of Stage IV data volumes. 

\subsubsection*{Modeling the galaxy population for informative priors}

If the likelihood tells us how a latent, i.e.~hidden true nature of a galaxy, maps into its observed-like realization, the prior tells us what kinds of latent galaxies the Universe actually produces. When interpreting the output of any photo-$z$ estimator according to Eqn.~\ref{eqn:bh}, there is no way to avoid making such prior assumptions (cf.~also the discussion of the "implicit prior"). Generative approaches tackle this problem by making the galaxy population model explicit. There are two broad ways of doing so, mirroring the phenotype/genotype language already introduced in the context of Eq. \ref{eqn:zbayes}.

In a phenotype-based view, the continuum of distinguishable galaxy types $t$ is defined in observational space. Galaxies are grouped into phenotypes or cells in a deep photometric manifold, and the inference problem becomes that of learning $p(z|t)$ and $p(t)$ across the manifold, and the transfer function between phenotypes and wide observations. This part of the chain can be equivalent to the clustering approaches discussed in Section~\ref{sec:1}. The model here is that galaxies come in a discrete set of observationally distinguishable classes. This approach remains close to the data and avoids the imprinting of biased color-redshift relations obtained from models that are not accurate enough.

In a genotype-based or physical view, one instead writes the prior with the help of a galaxy population model that describes the joint distribution of redshift evolving galaxy properties. Galaxy population models can be of the phenomenological or physical type. In phenomenological models \cite{Herbel2017,Tortorelli2020,Tortorelli2021,Amara2021,Moser2024,Fischbacher2025}, galaxies are described by their redshifts and noiseless photometric observables, such as positions, fluxes, and sizes. SEDs are constructed here as a combination of the same spectral templates $t$ that are used for photo-$z$ inference. Physical models, on the other hand, build a galaxy's description by its physical characteristics, such as stellar mass, star formation history, metallicity, dust attenuation, nebular component, and the presence of an AGN. These models make use of evolutionary stellar population synthesis
(hereafter, SPS, \cite{Conroy2013,Iyer2026}) to generate galaxy SEDs. SPS consists in exploiting the knowledge of stellar evolution to model the starlight emission of a galaxy. SPS is combined with models for the gas chemical evolution, dust attenuation and emission and AGN emission in generative SED codes (e.g. \textsc{FSPS} \cite{Conroy2010}, \textsc{CIGALE} \cite{Boquien2019}, \textsc{ProSpect} \cite{Robotham2020}) to  connect the physical properties of galaxies to their emitted light.

Physical models of galaxy populations for cosmological $N(z)$ inference and galaxy evolution studies include both the family of semi-analytical, hydro-dynamical and semi-empirical models (see \cite{Lapi2026} and references therein) and generative models designed for forward-modeling purposes. Examples of the latter category are \textsc{GalSBI} \cite{Fischbacher2025} and \textsc{GalSBI-SPS} \cite{Tortorelli2025}, \textsc{pop-cosmos} \cite{Thorp2025}, \textsc{Synthesizer} \cite{Lovell2025}, \textsc{DiffstarPop} \cite{Alarcon2025}, \textsc{PopSED} \cite{Li2024}, \textsc{PROVABGS} \cite{Hahn2023c} or \textsc{GalSyn} \cite{Abdurrouf2026}. Here the role of AI is more directly generative, with population models based on diffusion models (e.g. \textsc{pop-cosmos} \cite{Alsing2023}), normalizing flows (e.g. \textsc{PZFlow} \cite{Crenshaw2024}) or other flexible latent-density models \cite{Parker2025} that can represent the prior over the high-dimensional physical parameter space. AI is also used in the form of neural emulators that accelerates the otherwise expensive calculations conducted by generative SED codes based on SPS. These emulators map galaxy physical properties into either rest-frame noiseless SEDs (e.g. Speculator \cite{Alsing2020}) or observer and rest-frame magnitudes in any number of bands (e.g. \textsc{ProMage} \cite{Tortorelli2025b} and \textsc{Photulator} \cite{Thorp2024}), with more traditional MLPs sufficient to reach the precision requirements of photo-$z$ or cosmological $N(z)$ inference cases.

An important conceptual consequence follows. In discriminative AI, spectroscopy mainly appears as a training label for the inverse map. In generative AI, spectroscopy plays a different and in many ways more impactful role: it constrains the components of the generative model that broad-band photometry alone cannot pin down, such as emission line behavior, dust laws, metallicity evolution, or rare galaxy populations. These model components have been demonstrated in \cite{Tortorelli2024} to shift the inferred redshift distributions by amounts that are relevant for Stage IV cosmology. The issue is therefore no longer whether the spectroscopic sample is a representative training set for a black-box regressor, but how informative it is about the population model. At the same time, when properly accounting for the selection function in the hierarchical likelihood, spectroscopy of bright, low-redshift galaxies does inform at some model-dependent level the color-redshift relation of fainter and higher redshift galaxies. That shift of role and multiplication of utility of training data is one of the clearest distinctions between the discriminative and generative paradigms.

\subsubsection*{The inference process}

Once the prior and likelihood have been identified, one still has to perform the inference. This inference ideally is an inference of both the parameters defining the galaxy population, i.e. the "prior", and the properties of each individual galaxy in the sample. Large scale structure studies that require redshift distributions of galaxy samples almost exclusively depend upon constraining the galaxy population "prior" well. From a galaxy evolution perspective, there is a sense in which that prior itself is the main objective, since it describes what galaxies are like and how they evolve at any time.

In many realistic forward models, the inference process cannot be conducted by writing down an explicit analytic form for the likelihood given the complexity of the forward process or, even when that is possible, it can be too expensive to evaluate exactly. This is a situation where simulation-based inference (SBI, \cite{Cranmer2020}) is the method of choice. This is indeed the use case of the AI-aided forward model described in the previous sections.

SBI uses these forward simulations to learn or approximate either the joint posterior distribution of galaxy physical properties and population hyper-parameters or the likelihood of the model with which one then performs standard Monte Carlo Markov Chain (MCMC) inference of the same quantities. The most well known example of posterior estimation in the context of SBI is Approximate Bayesian Posterior (ABC, \cite{Beaumont2019}). In the simplest form of rejection ABC, this technique rejects or keeps candidate parameters $\theta_i$ drawn from the prior on the basis of a distance metric that encapsulates whether the simulated data are sufficiently close to the observed data. Several studies in galaxy evolution \cite{Cameron2012,Robin2014,Tortorelli2020,Tortorelli2021}, cosmology \cite{Weyant2013,Akeret2015,Alsing2019,Kacprzak2020,Fischbacher2025} and exoplanet research \cite{Kunimoto2020,Bryson2021} have successfully employed ABC to derive posterior estimates owning to its inherent conservative nature which mitigates the risk of convergence to local minima at the expense of approximated posteriors that are much broader than the "true" posterior. The sharp drop in acceptance rate as the distance threshold between data and simulation approaches zero \cite{Alsing2018,Cranmer2020} and its computationally intensive nature have led to the rise of AI driven SBI approaches based on neural density estimators \cite{Papamakarios2016,Lueckmann2017,Papamakarios2018,Alsing2018}. These make use of normalizing flows \cite{Papamakarios2017,Papamakarios2018,Papamakarios2019,Kobyzev2021} or mixture density networks \cite{Bishop1994} to derive both individual galaxy \cite{Harvey2026} and population posteriors \cite{Ho2024,Lovell2025b} which in turn enable ensemble redshift distributions estimates and cosmological analyses \cite{Hahn2023b,Wietersheim-Kramsta2025}. These techniques are capable of modeling complex posteriors using fewer, amortized simulations, but may yield overconfident results, particularly when sample sizes are small or priors are mis-specified \cite{Hermans2021,Tam2022}.

Amortization is an aspect that is going to be particularly compelling for the Rubin and Euclid era. A fully faithful generative posterior for each galaxy, conditioned on a population prior and realistic data model, is too expensive to obtain from scratch billions of times. One therefore expects a hybrid pattern to emerge: expensive forward simulations and hierarchical inference are used offline to train emulators or posterior estimators, which can then be deployed at scale. In this sense, generative AI is an efficient strategy to scale Bayesian photo-$z$ inference to Stage IV surveys data volumes.

\subsubsection*{Pros and Cons of generative AI in photo-$z$}

Generative AI and forward models help in directly addressing the failure modes identified in Section \ref{sec:generative}. They replace implicit priors with explicit, learnable population models. They replace dependence on spectroscopic labels alone with joint constraints that properly use heterogeneous datasets. And they upgrade the likelihood from a noise approximation to a simulation-based explicit model of measurement, blending, and selection. This changes the target of the inference process from per-object regression to calibrated posterior inference on $N(z)$, galaxy population parameters, and eventually cosmological parameters themselves. This is particularly compelling for weak lensing and galaxy clustering, where the dominant requirement is accuracy on the redshift distribution rather than a minimal redshift uncertainty for individual galaxies.

Generative AI is not the solution to everything and should be challenged like any other model in science: (i) a flexible prior is still a prior and can still be mis-specified; extensions need to be evaluated, but also there is an intrinsic requirement to curb complexity where it becomes unfeasible to inform with data; (ii) an emulator can still fail in corners of parameter space that matter disproportionately for cosmology; (iii) a simulator can still omit relevant observational systematics; (iv) an amortized posterior estimator can still be mis-calibrated even when its point predictions look excellent. For that reason, validation in this framework must go beyond the usual photo-$z$ scatter and outlier metrics. Coverage tests, posterior predictive checks of known physical relations between galaxies, model mis-specification tests, sensitivity to spectroscopic incompleteness, and end-to-end tests of cosmological parameter shifts under controlled substitutions all become central. 

\subsubsection*{AI and photo-$z$}

The resulting picture is that in photo-$z$s there is a confluence of expertise in galaxy evolution, cosmology, statistics, and AI techniques. The prior term in the Bayesian equation equates a model that describes galaxy evolution, the likelihood becomes a model of how surveys observe and measure galaxies. Discriminative AI has been developed and used in a diversity of ways to approximate this inference. Generative AI, on the contrary, provides flexible density estimators, emulators, and inference engines that have the potential to make principled inference scalable to Stage IV requirements and data volumes.

\bibliographystyle{JHEP} 
\bibliography{bibliography}

\end{document}